\documentclass{aa}

\usepackage{graphicx}
\usepackage{txfonts}
\usepackage{subcaption}
\usepackage{booktabs}
\usepackage{xcolor}

\usepackage[colorlinks=true,citecolor=blue,linkcolor=blue]{hyperref}

\usepackage{tikz}
\usetikzlibrary{decorations.pathmorphing,decorations.markings,calc,patterns,arrows.meta,positioning}

\DeclareMathOperator{\Rr}{Re}
\DeclareMathOperator{\Ii}{Im}

\begin{document}

\title{A user-friendly package and workflow for generating effective homogeneous rheologies for the study of the long-term orbital evolution of multilayered planetary bodies}

\titlerunning{A user-friendly package for generating effective homogeneous rheologies for multilayered planetary bodies}

\author{Yeva Gevorgyan\inst{1}}
\authorrunning{Y. Gevorgyan}
\institute{King Abdullah University of Science and Technology (KAUST), Thuwal, Saudi Arabia \\ \email{yeva.gevorgyan@kaust.edu.sa}}

\date{Received March 14, 2026; accepted April 30, 2026}

\abstract
{Tidal dissipation plays an important role in the long-term orbital evolution of planets and moons. For stratified viscoelastic bodies, evaluating the frequency-dependent tidal response during long-term dynamical integrations is computationally expensive, since it requires repeatedly solving the internal deformation problem. Therefore, orbital evolution codes benefit from effective homogeneous rheologies that approximate the dissipation of stratified interiors at low computational cost.}
{We present a user-friendly, open-source Wolfram Language package that constructs an effective homogeneous generalized Voigt rheology for a spherically symmetric, incompressible layered body with Maxwell solid layers. The package converts the tidal response of a layered interior model into the parameters required by time-domain simulations of tidal evolution.}
{The package combines three components: (i) a forward computation of the degree-2 tidal Love number based on the propagator-matrix formulation for incompressible stratified viscoelastic bodies; (ii) numerical identification of the secular relaxation poles and residues of the layered model; and (iii) inversion of the resulting response into the compliance of an equivalent homogeneous generalized Voigt body. The implementation is based on the equivalence established for multilayer Maxwell bodies and includes an optional dominant-mode selection procedure for obtaining reduced rheological models over a prescribed frequency range.}
{The package returns the parameters of the equivalent homogeneous model, including elastic, gravitational, viscous, and Voigt-element contributions, in a format that can be used by external tidal-evolution codes. As a case study, we apply the package to a five-layer lunar interior model and obtain its equivalent generalized Voigt representation, together with a reduced model that preserves the tidal response over the frequency interval relevant for orbital evolution while using fewer relaxation elements.}
{The package provides a reproducible implementation of the reduction from stratified viscoelastic interiors to effective homogeneous rheologies. It allows tidal dissipation from layered models to be included in long-term orbital and spin-evolution calculations without solving the full layered boundary-value problem at each step.}


\maketitle
\nolinenumbers

\section{Introduction}
\label{sec:int}

Tidal deformation and energy dissipation play an important role in the long-term orbital evolution of planets and moons. The magnitude and frequency of tidal dissipation depend on the internal structure of the deformable body, in particular on the distribution of density, elasticity, and viscosity.

Tidal deformation is modeled with the frequency-dependent degree-2 tidal Love number $k_2(s)$, that can be obtained by solving the governing equations for a radially stratified viscoelastic body. Here, the stratified forward model follows the propagator-matrix formulation for incompressible, spherically symmetric stratified bodies in \citet[Chapter 2]{sabadini2016global}. Solid layers are modeled with Maxwell rheology, and liquid layers through the appropriate small-shear limit. This approach is a consistent way to evaluate the tidal response of the body, but can become too costly to solve repeatedly within long-term orbital or spin-evolution integrations.

Long-term dynamical evolution codes often rely on simplified homogeneous rheologies or empirical prescriptions, such as constant time lag, constant phase lag, or single-element Maxwell models. These approximations are computationally convenient, but generally fail to reproduce the multiple relaxation timescales and dissipation peaks that arise in layered bodies \citet{Gev2021, 2020A&A...644A.165B}. In \citet{Gev2023}, the authors showed that, under the assumptions of a simple multilayer rheology, the tidal response of a stratified body can be modeled by a homogeneous body with a sufficiently complex generalized rheology, like a generalized Voigt model. This motivated us to develop a relatively simple workflow that allows one to replace the stratified problem by an effective homogeneous model that preserves the frequency-dependent tidal response.

In this work we present an open-source Wolfram Language package\footnote{\href{https://github.com/ygevorgyan/HomogeneousModel}{https://github.com/ygevorgyan/HomogeneousModel}} that takes a user-specified multilayered model and constructs the corresponding effective homogeneous generalized Voigt rheology. The input file contains the layer radii, densities, shear moduli, viscosities, and forcing frequency, and the main routine performs the full reduction pipeline: computation of $k_2(s)$ for the layered model, identification of secular poles and residues, inversion to the equivalent compliance function, and extraction of the generalized Voigt parameters. An optional reduction step identifies dominant relaxation modes over a user-defined frequency interval, yielding a compact model when an exact full representation is unnecessary.

Our implementation was partly motivated by the need to connect detailed geophysical interior models with modern time-domain orbital integrators such as the RheoVolution \citep{2025A&A...693A...5D,2025ascl.soft08007D}, which can evolve deformable bodies with complex homogeneous rheologies but require as input the parameters of an effective rheological model rather than a full multilayer interior description. The package converts the tidal response of a stratified viscoelastic body into the parameters of a homogeneous generalized Voigt model that can be used in such simulations.

The paper is organized as follows. In Section~2, we summarize the theoretical framework for the effective homogeneous rheology and its relation to the layered tidal response. In Section~3, we describe the structure of the software package, its inputs and outputs, and the computational workflow. In Section 4, we illustrate the method with an application to a five-layer model of the Moon and show how the package recovers the equivalent generalized Voigt rheology and, if desired, an optional reduced model that captures the tidal response over the frequency interval of interest. In Section~5 we conclude the paper with observations on the purpose and scope of the work and a few perspectives for improvement.

\section{Theoretical background}
\label{sec:theory}

This section summarizes the theoretical elements relevant to the software pipeline. The forward computation of the layered tidal response follows the propagator-matrix formulation for incompressible, stratified viscoelastic bodies presented in \citet{sabadini2016global}, while the reduction to an equivalent homogeneous rheology is based on \citet{Gev2023}.

\subsection{Layered viscoelastic model}

Consider a spherically symmetric body of $N$ incompressible, homogeneous layers, each with outer radius $r_k$, density $\rho_k$, rigidity $\mu_k$, and viscosity $\eta_k$. Solid layers are assumed to have Maxwell rheology, and their complex shear modulus in the Laplace domain is
\begin{equation}
\hat{\mu}_k(s)=\frac{\mu_k \eta_k s}{\mu_k+\eta_k s}.
\end{equation}
The tidal response is described by the degree-2 Love number $k_2(s)$, obtained from the propagator-matrix solution of the governing equations for a stratified body. All layers are treated within a unified propagator formulation, and liquid layers are handled in the small-shear limit, see Subsection~\ref{sec:workflow}.

For a finite multilayer Maxwell body, $k_2(s)$ is a rational function of the Laplace variable and admits the mode decomposition \citep[Eq.~26]{Gev2023}
\begin{equation}
k_2(s)=k_E+\sum_{i=1}^{n+1}\frac{r_i}{s-s_i},
\label{eq:k2_partial}
\end{equation}
where $k_E$ is the elastic Love number, $s_i<0$ are the secular poles, $r_i$ are the corresponding residues, and $\tau_i=-s_{i}^{-1}$ are the relaxation times. Here $n$ denotes the number of Kelvin--Voigt elements of the equivalent homogeneous model, so the layered body has $n+1$ normal modes, in agreement with \citet[Eq.~22]{Gev2023}. Thus, the layered tidal response is fully characterized by the set of parameters $\{k_E,s_i,r_i\}$.

\subsection{The homogeneous model}

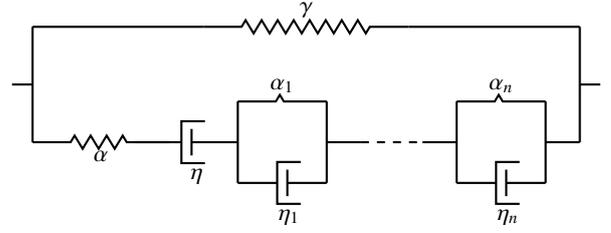
\begin{figure}[ht]
\begin{center}
\begin{tikzpicture}[scale=0.9, transform shape]
\tikzstyle{spring}=[thick, decorate, decoration={zigzag, pre length=0.5cm, post length=0.5cm, segment length=6}]
\tikzstyle{damper}=[thick, decoration={markings,
  mark connection node=dmp,
  mark=at position 0.5 with
  {
    \node (dmp) [thick, inner sep=0pt, transform shape, rotate=-90, minimum width=15pt, minimum height=3pt, draw=none] {};
    \draw [thick] ($(dmp.north east)+(5pt,0)$) -- (dmp.south east) -- (dmp.south west) -- ($(dmp.north west)+(5pt,0)$);
    \draw [thick] ($(dmp.north)+(0,-5pt)$) -- ($(dmp.north)+(0,5pt)$);
  }
}, decorate]
\tikzstyle{ground}=[fill,pattern=north east lines,draw=none,minimum width=0.75cm,minimum height=0.3cm]

            \draw [thick] (0,0.8) --  (0,2.5);

           \draw [spring] (0,0.8) -- node[below] {$\alpha$} (2,0.8);
            \draw [damper] (1.5,0.8) -- (3,0.8);
            \node at (2.4,0.3) {$\eta$};

            \draw [thick] (3,0.2) --  (3,1.4);

           \draw [spring] (3,1.4) -- node[above] {$\alpha_1$} (4.3,1.4);
            \draw [damper] (3,0.2) -- (4.3,0.2);
            \node at (3.75,-0.3) {$\eta_{1}$};

            \draw [thick] (4.3,0.2) --  (4.3,1.4);

            \draw [thick] (4.3,0.8) --  (4.8,0.8);
            \draw [thick,dashed] (4.8,0.8) -- (5.7,0.8);
            \draw [thick] (5.7,0.8) -- (6.2,0.8);

            \draw [thick] (6.2,0.2) --  (6.2,1.4);

           \draw [spring] (6.2,1.4) -- node[above] {$\alpha_n$} (7.5,1.4);
            \draw [damper] (6.2,0.2) -- (7.5,0.2);
            \node at (6.95,-0.3) {$\eta_{n}$};

            \draw [thick] (7.5,0.2) --  (7.5,1.4);

            \draw [thick] (7.5,0.8) --  (8,0.8);

            \draw [thick] (8,0.8) --  (8,2.5);

            \draw [thick] (0,2.5) --  (2.5,2.5);
           \draw [spring] (2.5,2.5) -- node[above] {$\gamma$} (5.5,2.5);
            \draw [thick] (5.5,2.5) --  (8,2.5);

            \draw [thick] (-0.3,1.65) --  (0,1.65);
            \draw [thick] (8,1.65) --  (8.3,1.65);

      \end{tikzpicture}
\end{center}
\caption{Schematic representation of the generalized Voigt model.}
\label{fig:voig-osc}
\end{figure}

The equivalent homogeneous representation of stratified bodies is given by a generalized Voigt rheology in Figure \ref{fig:voig-osc}. The Love number of the body is \citep[Eq.~2]{Gev2023}
\begin{equation}
k_2(s)=\frac{3\mathrm{I}_\circ G}{R^5}\,C(s),
\label{eq:k2C_short}
\end{equation}
where $\mathrm{I}_\circ$ is the mean moment of inertia, $C(s)$ is the complex compliance, and $R$ is the mean radius of the body. The gravitational rigidity $\gamma$ is defined from the fluid Love number $k_f$ as \citep[Eq.~4]{Gev2023}
\begin{equation}
\gamma=\frac{3\mathrm{I}_\circ G}{R^5}\frac{1}{k_f}.
\label{eq:gamma_short}
\end{equation}
For the generalized Voigt rheology, the complex compliance is decomposed as \citep[Eq.~14]{Gev2023}
\begin{equation}
C(s)=\frac{1}{\alpha+\gamma}+C_v(s),
\label{eq:Csplit_short}
\end{equation}
where $\alpha$ is the elastic rigidity and $C_v(s)$ contains the viscous contribution of the dashpot and a finite number of Kelvin--Voigt elements in Figure \ref{fig:voig-osc}. The equivalent homogeneous model is specified by the parameters $\{\gamma, \alpha, \eta, \alpha_i, \eta_i\}$. In these expressions $\gamma$, $\alpha$, and $\alpha_i$ have dimensions of $\mathrm{s^{-2}}$, as does the mapping constant $\Theta=3\mathrm{I}_\circ G/R^5$, while $\eta$ and $\eta_i$ have dimensions of $\mathrm{s^{-1}}$. To convert to SI units of Pa and $\mathrm{Pa\,s}$, we divide by the rescaling factor $\frac{152\pi}{15}\frac{R}{m_1}$ \citep[Section 7.1.1]{gev2020}, where $m_1$ is the mass of the body. The outputs reported by the package (for example in Table~\ref{tab:effective_parameters}) are given in these units.

\subsection{Extraction of the homogeneous model parameters}
\label{sec:extraction}

The parameters of the equivalent homogeneous generalized Voigt rheology are obtained from the layered pole--residue data $\{k_E, s_i, r_i\}_{i=1}^{n+1}$ by a purely algebraic procedure. Writing the layered compliance
\begin{equation}
C_L(s)=\frac{R^5}{3\mathrm{I}_\circ G}\,k_2(s)
=k_E^C+\sum_{i=1}^{n+1}\frac{r_i^C}{s-s_i},
\label{eq:CL}
\end{equation}
where $k_E^C$ and $r_i^C$ are $k_E$ and $r_i$ rescaled by $R^5/(3\mathrm{I}_\circ G)$, the elastic and gravitational parameters follow from the asymptotic limits of the compliance,
\begin{equation}
\gamma=\left(k_E^C-\sum_{i=1}^{n+1}\frac{r_i^C}{s_i}\right)^{-1},
\qquad
\alpha=\frac{1}{k_E^C}-\gamma.
\label{eq:gamma_alpha}
\end{equation}
The remaining parameters are $\eta$ and the $n$ Kelvin--Voigt pairs $\{(\alpha_i,\eta_i)\}_{i=1}^{n}$, totalling $2n+1$ unknowns. Attempting to extract these parameters from the $n+1$ pole conditions of the generalized Voigt model would be both underdetermined and nonlinear. Instead, we construct the complex compliance of the visco-elastic element $\hat{J}(s)$ \citep[Eq.~13]{Gev2023},
\begin{equation}
\hat{J}(s)=\frac{1}{\alpha}+\frac{1}{\eta s}+\sum_{i=1}^{n}\frac{1}{\alpha_i+s\eta_i},
\label{eq:Jhat}
\end{equation}
directly from the layered data through the relation $C^{-1}(s)=\gamma+\hat{J}^{-1}(s)$, which yields
\begin{equation}
\hat{J}(s)=\frac{C_L(s)}{1-\gamma\,C_L(s)}.
\label{eq:Jhat_from_CL}
\end{equation}
All Kelvin--Voigt parameters are then read off from the partial-fraction decomposition of $\hat J(s)$.

Writing the layered compliance \eqref{eq:CL} as a single fraction $C_L(s)=N(s)/D(s)$, where $D(s)=\prod_{i=1}^{n+1}(s-s_i)$ and $N(s)$ is a polynomial of degree $n+1$, equation \eqref{eq:Jhat_from_CL} gives
\begin{equation}
\hat{J}(s)=\frac{N(s)}{D(s)-\gamma\,N(s)}=\frac{N(s)}{s\,Q(s)},
\label{eq:Jhat_rational}
\end{equation}
where the denominator vanishes at $s=0$ as a direct consequence of the definition of $\gamma$ in \eqref{eq:gamma_alpha}, and $Q(s)$ is a polynomial of degree $n$. The poles of $\hat{J}(s)$ are therefore $s=0$ and the $n$ roots of $Q(s)$, which are real, negative, and simple \citep[from the interlacing property in Eq.~22 of][]{Gev2023}.

The residue of $\hat{J}(s)$ at $s=0$ comes entirely from the dashpot term in \eqref{eq:Jhat}, the only singular contribution at the origin. Matching the residue computed from \eqref{eq:Jhat_rational} gives
\begin{equation}
\eta=\frac{Q(0)}{N(0)}.
\label{eq:eta_extract}
\end{equation}
At each remaining pole $q_i$, only the $i$-th Kelvin--Voigt term is singular, and writing $1/(\alpha_i+s\eta_i)=\eta_i^{-1}/(s-q_i)$ with $q_i=-\alpha_i/\eta_i$, residue matching yields
\begin{equation}
\eta_i=\frac{q_i\,Q'(q_i)}{N(q_i)},
\qquad
\alpha_i=-q_i\eta_i=\frac{-q_i^2\,Q'(q_i)}{N(q_i)}.
\label{eq:KV_extract}
\end{equation}

No nonlinear system is solved. The only numerical step is the polynomial root-finding of $Q(s)$, a standard operation for a polynomial whose roots are real, negative, and simple. For a layered body with $n+1$ normal modes, the $2n+3$ data points $\{k_E^C, r_i^C, s_i\}$ uniquely determine the $2n+3$ parameters $\{\gamma,\alpha,\eta,\alpha_i,\eta_i\}$ of the equivalent homogeneous model.

\section{Software implementation and workflow}

The package is written in Wolfram Language as a modular pipeline for constructing the homogeneous generalized Voigt rheology of a layered viscoelastic body. The model is not hardcoded to any specific planetary interior: all physical and numerical inputs are supplied by the user through a configuration file, and the same workflow applies to any incompressible, spherically symmetric multilayer model satisfying the assumptions discussed in Section~\ref{sec:theory}.

\subsection{Package structure}
The code is organized into a small set of modules with distinct roles. The file \texttt{ModelInput.wl} contains the user-specified physical model and numerical settings. The core routines are called by \texttt{computeHomogeneous}, implemented in \texttt{HomogeneousRheology.wl}, which evaluates the layered tidal response, extracts the secular poles and residues, and returns the parameters of the equivalent generalized Voigt model. A driver script, \texttt{RunHomogeneous.wl}, loads the modules, executes the full pipeline, prints summary tables, and generates comparison plots. A separate script, \texttt{VisualizeLayers.wl}, produces a cross-sectional visualization of the layered body directly from the same input file.

\subsection{Input configuration}

The file \texttt{ModelInput.wl} contains both the physical description of the layered body and a small set of numerical controls for the inversion pipeline. The same workflow can be applied to any spherically symmetric multilayer configuration once the corresponding layer properties and forcing frequency are provided.

The physical model is specified through the arrays of outer radii, densities, rigidities, and viscosities of the layers, together with the tidal forcing frequency. An optional list of layer names may be supplied for visualization purposes. All physical quantities are given in SI units, and all arrays must have the same length, which determines the number of layers in the model. Liquid layers are identified through the condition $0 < \mu < 1$~Pa; for such layers, the method uses an effective elasticity value to regularize the small-shear limit \citep{sabadini2016global}.

The same input file contains numerical settings that control the precision of the calculations, the search interval used to detect the secular poles, and the frequency range and threshold adopted in the dominant-mode selection step. These parameters are not part of the planetary model but determine how the software resolves the relaxation spectrum and how aggressively the final generalized Voigt model is reduced. Their default values are listed in Table~\ref{tab:package_inputs}.

The single tidal forcing frequency $\omega$ and the frequency interval $[\omega_{\min},\omega_{\max}]$ play distinct roles. The rheological parameters of the equivalent homogeneous model are frequency-independent, determined entirely by the secular poles and residues of the layered $k_2(s)$. The single $\omega$ is used only to evaluate and report $k_2(i\omega)$ at the frequency of interest for the body under consideration. For the Moon this corresponds to the orbital frequency $\omega = 2.662 \times 10^{-6}~\mathrm{rad\,s^{-1}}$, associated with the 27.32-day orbital period, which we use throughout the case study in Section~\ref{sec:casestudy}. The interval $[\omega_{\min},\omega_{\max}]$ enters only in the optional dominant-mode reduction step, where it defines the band over which each Kelvin--Voigt element is assessed for retention in the reduced model.

\begin{table}[htb]
\caption{Main user inputs and numerical controls of the Wolfram package.}
\label{tab:package_inputs}
\centering
\footnotesize
\setlength{\tabcolsep}{5pt}
\renewcommand{\arraystretch}{1.15}
\begin{tabular}{p{0.32\columnwidth}p{0.18\columnwidth}p{0.38\columnwidth}}
\toprule
Key & Units/default & Description \\
\midrule
\multicolumn{3}{l}{Physical model (\texttt{\$ModelConfig})} \\[3pt]
\texttt{rk} & m & Layers' outer radii \\
\texttt{rhok} & kg\,m$^{-3}$ & Layer densities \\
\texttt{muk} & Pa & Layers' shear moduli \\
\texttt{etak} & Pa\,s & Layers' viscosities \\
\texttt{omega} & rad\,s$^{-1}$ & Tidal forcing frequency \\
\texttt{layerNames} & optional & Optional list of layer labels \\
\midrule
\multicolumn{3}{l}{Numerical settings (\texttt{\$AnalysisConfig})} \\[3pt]
\texttt{workingPrecision} & 50 & Numerical working precision in digits \\
\texttt{gridLogMin} & $-20$ & Lower bound of the pole-search interval in $\log_{10}|s|$ \\
\texttt{gridLogMax} & $-3$ & Upper bound of the pole-search interval in $\log_{10}|s|$ \\
\texttt{gridPoints} & 2000 & Number of grid points used to detect sign changes in the secular determinant \\
\texttt{omegaInterestMin} & $10^{-10}$ rad\,s$^{-1}$ & Lower bound of the frequency interval used for dominant-mode selection \\
\texttt{omegaInterestMax} & $10^{-5}$ rad\,s$^{-1}$ & Upper bound of the frequency interval used for dominant-mode selection \\
\texttt{dominantThreshold} & $1/50$ & Fractional contribution threshold used to retain dominant Voigt elements \\
\bottomrule
\end{tabular}
\tablefoot{Physical quantities are supplied through \texttt{ModelInput.wl}; optional numerical settings control pole search and dominant-mode selection.}
\end{table}

\subsection{Main routine and computational workflow}
\label{sec:workflow}

\begin{figure}[htbp]
\begin{center}
\begin{tikzpicture}[
  box/.style={draw, rounded corners, minimum width=8.5cm,
              minimum height=0.9cm, align=center, font=\small},
  arrow/.style={-{Stealth[length=2.5mm]}, thick},
  lbl/.style={font=\footnotesize\itshape, text=black!60},
  node distance=0.4cm
]

\node[box, fill=blue!8] (input)
  {Layer parameters $(r_k, \rho_k, \mu_k, \eta_k)$ and tidal frequency $\omega$};

\node[box, fill=orange!10, below=of input] (k2)
  {Compute $k_2(s)$ via propagator matrix};

\node[box, fill=orange!10, below=of k2] (poles)
  {Find secular poles $\{s_j\}$};

\node[box, fill=orange!10, below=of poles] (res)
  {Compute residues $\{r_j\}$ and $k_E$, $k_f$};

\node[box, fill=green!10, below=of res] (J)
  {Build $\hat{J}(s)$ as rational function};

\node[box, fill=green!10, below=of J] (voigt)
  {Extract Voigt parameters $(\gamma, \alpha, \eta, \alpha_i, \eta_i, \tau_i)$};

\node[box, fill=violet!10, below=of voigt] (dom)
  {Identify dominant modes $> f_{\mathrm{threshold}}$};

\node[box, fill=blue!8, below=of dom] (output)
  {Equivalent homogeneous generalized Voigt model};

\draw[arrow] (input) -- (k2);
\draw[arrow] (k2) -- (poles);
\draw[arrow] (poles) -- (res);
\draw[arrow] (res) -- (J);
\draw[arrow] (J) -- (voigt);
\draw[arrow] (voigt) -- (dom);
\draw[arrow] (dom) -- (output);

\node[lbl, rotate=90, anchor=south] at ($(k2.north west)!0.5!(res.south west) + (-0.3,0)$)
  {Forward};
\draw[thick, black!30] ($(k2.north west)+(-0.2,0.1)$)
  -- ($(res.south west)+(-0.2,-0.1)$);

\node[lbl, rotate=90, anchor=south] at ($(J.north west)!0.5!(voigt.south west) + (-0.3,0)$)
  {Inversion};
\draw[thick, black!30] ($(J.north west)+(-0.2,0.1)$)
  -- ($(voigt.south west)+(-0.2,-0.1)$);

\node[lbl, rotate=90, anchor=south] at ($(dom.north west)!0.5!(dom.south west) + (-0.3,0)$)
  {Reduction};
\draw[thick, black!30] ($(dom.north west)+(-0.2,0.1)$)
  -- ($(dom.south west)+(-0.2,-0.1)$);

\end{tikzpicture}
\end{center}
\caption{Workflow of the \texttt{computeHomogeneous} pipeline.}
\label{fig:workflow}
\end{figure}
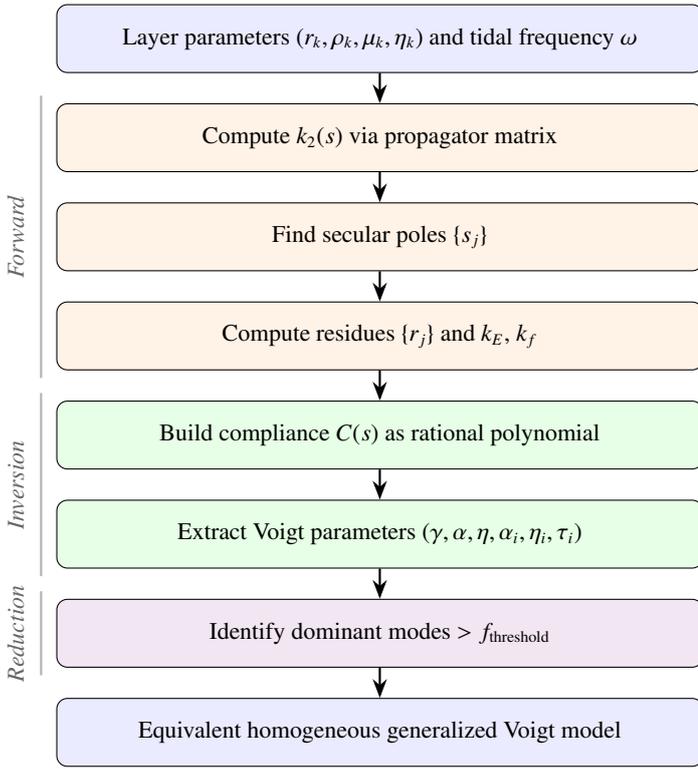

The main routine
\begin{center}
\texttt{computeHomogeneous[rk,rhok,muk,etak,omega,opts]}
\end{center}
performs the parameter extraction through the sequence of operations illustrated in Fig.~\ref{fig:workflow}. It follows the same three-step structure of Sections \ref{sec:int} and \ref{sec:theory}: forward computation of the layered Love number, mode decomposition, and rheological inversion. The code first evaluates the elastic Love number $k_E$ and the complex tidal response $k_2(i\omega)$ for the layered body. It then searches for the zeros of the secular determinant on the negative real axis, refines them numerically, and computes the corresponding residues. From these quantities it evaluates the fluid Love number $k_f$ and the mapping constant $\Theta=3\mathrm{I}_\circ G/R^5$, and applies the algebraic procedure of Section~\ref{sec:extraction} to extract the parameters $\{\gamma, \alpha, \eta, \alpha_i, \eta_i\}$ of the equivalent generalized Voigt model.

The forward layered problem is evaluated through a unified $6\times6$ propagator matrix approach for all layers. Liquid layers are treated by small-shear regularization, such that solid and liquid regions are handled within the same numerical framework. The current package does not switch to a separate Clairaut propagator for liquids, but instead uses a single propagation formalism for arbitrary combinations of solid and liquid layers. A true liquid has $\mu = 0$, but the $Y$-matrix \citep[Eq.~2.42]{sabadini2016global} contains terms of the form $\mu/r$ in rows 3, 4, and 6. Setting $\mu = 0$ exactly would therefore make these rows degenerate, causing the linear solver to fail when computing the propagator
\begin{equation}
P(r_2,r_1) = Y(r_2)\,Y^{-1}(r_1).
\end{equation}
Instead, we assign a very small effective rigidity $\hat{\mu} \approx 10^{-20}\,\mathrm{Pa}$ to liquid layers. This keeps the matrix full-rank and invertible, while making the shear-stress rows effectively zero to machine precision. This value is negligible compared with physical shear moduli, which are typically of order $10^{9}$--$10^{11}\,\mathrm{Pa}$, but remains large enough to preserve numerical stability at all reasonable (actually, even at quite stringent) working precisions. This is a standard regularization in viscoelastic normal-mode calculations that avoids introducing a separate $2\times2$ or $4\times4$ propagator for liquid layers while yielding results indistinguishable from the inviscid limit to within working precision.

After the full generalized Voigt model is obtained, the package performs an optional dominant-mode reduction, following the criterion introduced in \citet[Appendix~A]{Gev2023}. For each Voigt element, it evaluates the maximum fractional contribution to $-\Ii[C(i\omega)]$ over a user-defined frequency interval $[\omega_{\min},\omega_{\max}]$. Elements whose contribution exceeds the prescribed threshold are kept in the reduced model. This produces a compact, effective rheology targeted to the frequency interval most relevant for the intended dynamical application.

\subsection{Output}

The main routine \texttt{computeHomogeneous} returns the layered response and its homogeneous representation as a Wolfram \texttt{Association}, and also exports the corresponding results to a JSON file. The package also returns the intermediate modes used in the inversion. In this way, the user has access both to the physical outputs of interest--- the effective generalized Voigt parameters---and to the pole--residue decomposition that connects the layered model to the homogeneous model. The returned quantities are summarized in Table~\ref{tab:package_outputs}.

The output falls into three groups: the Love numbers of the layered body, the constants used in the layered-to-homogeneous mapping, and the relaxation modes and rheological parameters of the equivalent generalized Voigt model. This organization separates diagnostic quantities from the final effective model. The Love-number and mode outputs are used to check the inversion, and the generalized Voigt parameters are used in downstream applications. The package also generates summary tables, dominant-mode diagnostics, and comparison plots. The user can inspect the mode decomposition, compare the full and reduced models, or export the reduced generalized Voigt parameters to an external integrator such as RheoVolution \citep{2025A&A...693A...5D,2025ascl.soft08007D}.

\begin{table}[htb]
\caption{Quantities returned by \texttt{computeHomogeneous}.}
\label{tab:package_outputs}
\centering
\footnotesize
\setlength{\tabcolsep}{5pt}
\renewcommand{\arraystretch}{1.15}
\begin{tabular}{p{0.20\columnwidth}p{0.22\columnwidth}p{0.46\columnwidth}}
\toprule
Key & Units & Description \\
\midrule
\texttt{kE} & dimensionless & Elastic Love number $k_E$ \\
\texttt{kf} & dimensionless & Fluid Love number $k_f$ \\
\texttt{k2Re} & dimensionless & Real part of $k_2(i\omega)$ \\
\texttt{k2Im} & dimensionless & Imaginary part of $k_2(i\omega)$ \\
\texttt{gamma} & Pa & Gravitational rigidity $\gamma$ \\
\texttt{alpha} & Pa & Elastic rigidity $\alpha$ \\
\texttt{eta} & Pa\,s & Viscosity of the isolated dashpot \\
\texttt{prefactor} & s$^{-2}$ & Mapping prefactor $\Theta = 3\mathrm{I}_\circ G/R^5$ \\
\texttt{Fconv} & Pa\,s$^2$ & Conversion factor to physical units \\
\texttt{nPoles} & integer & Number of secular poles (equal to $n+1$ in the notation of Section~\ref{sec:extraction}) \\
\texttt{poles} & s$^{-1}$ & Secular pole locations $\{s_j\}$ \\
\texttt{residues} & dimensionless & Residues $\{r_j\}$ in the mode decomposition of $k_2(s)$ \\
\texttt{nVoigt} & integer & Number of Kelvin--Voigt elements (equal to $n$ in the notation of Section~\ref{sec:extraction}) \\
\texttt{tauV} & s & Relaxation times $\{\tau_i\}$ of the Voigt elements \\
\texttt{alphaV} & Pa & Spring rigidities $\{\alpha_i\}$ of the Voigt elements \\
\texttt{etaV} & Pa\,s & Dashpot viscosities $\{\eta_i\}$ of the Voigt elements \\
\bottomrule
\end{tabular}
\tablefoot{The routine returns both the intermediate mode information of the layered model and the parameters of the equivalent homogeneous generalized Voigt rheology.}
\end{table}

\section{Case study: A five-layer model of the Moon}
\label{sec:casestudy}

\begin{figure}[ht]
    \centering
    \includegraphics[width=\columnwidth]{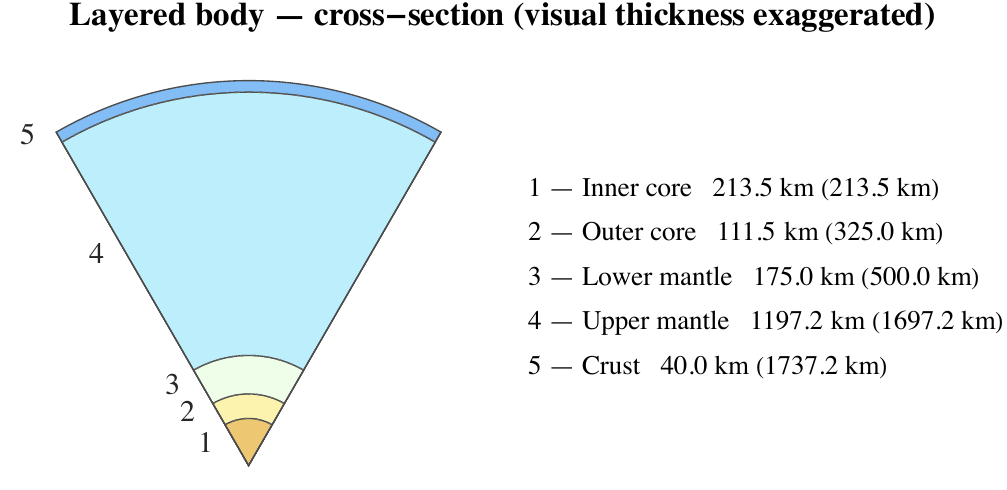}
    \caption{Cross-section depiction of the 5-layer lunar interior model automatically generated by the \texttt{VisualizeLayers.wl} routine. The numbered list indicates the thickness of each corresponding layer, along with the position (radius) of its outer boundary in parentheses.}
    \label{fig:lunar_cross_section}
\end{figure}

To test the package, we use the five-layer lunar model of \citet{Gev2023} to illustrate the equivalence between a multilayer Maxwell body and an effective homogeneous rheology. The model parameters are summarized in Table~\ref{tab:lunar_model}, and its cross-section is illustrated in Fig.~\ref{fig:lunar_cross_section}; the figure was automatically produced by the \texttt{VisualizeLayers.wl} module. The Moon is the default example configuration in the package. The package was also tested for Io, Europa, and Enceladus using models and data from \cite{2004Icar..171..391H}, \cite{2006JGRE..11112005W}, \cite{2016JGRE..121.2211B}, and \cite{Thomas2016}, although we are only going to report the results for the Moon. The \texttt{ModelInput} files for these additional test cases can be found in the software package repository.

\begin{table}[b]
\centering
\caption{Physical parameters for the 5-layer viscoelastic model of the lunar interior.}
\label{tab:lunar_model}
\resizebox{\columnwidth}{!}{%
\begin{tabular}{@{}lcccc@{}}
\toprule
Layer & $r_k$ (km) & $\rho_k$ (kg/m$^3$) & $\mu_k$ (GPa) & $\eta_k$ (Pa$\cdot$s) \\ \midrule
1. Inner Core & 213.52 & 7720.16 & 40 & $10^{21}$ \\
2. Outer Core & 325.00 & 6700.00 & $10^{-10}$ & 1 \\
3. Lower Mantle & 500.00 & 3800.00 & 60 & $5 \times 10^{16}$ \\
4. Upper Mantle & 1697.15 & 3356.00 & 62.5 & $10^{21}$ \\
5. Crust & 1737.15 & 2735.00 & 15 & $10^{23}$ \\ \bottomrule
\end{tabular}%
}
\tablefoot{Values taken from \citet[Table~2]{Matsuyama2016} and \citet{Gev2023}.}
\end{table}

\subsection{The homogeneous model}

The software extracts the secular relaxation poles and residues listed in Table~\ref{tab:secular_modes}. These modes provide the normal-mode decomposition of the lunar tidal response and are used to construct the equivalent homogeneous rheology. The corresponding inversion yields the full equivalent generalized Voigt model, whose parameters are listed in Table~\ref{tab:voigt_elements} and \ref{tab:effective_parameters}.

\begin{table}[htbp]
\centering
\caption{Secular relaxation modes extracted for the 5-layer lunar model.}
\label{tab:secular_modes}
\begin{tabular}{@{}cccc@{}}
\toprule
$j$ & Pole $s_j$ (s$^{-1}$) & $\tau_j$ (s) & Residue $r_j$ \\ \midrule
1 & $-9.459 \times 10^{-7}$ & $1.057 \times 10^{6}$ & $8.109 \times 10^{-10}$ \\
2 & $-6.342 \times 10^{-7}$ & $1.577 \times 10^{6}$ & $8.531 \times 10^{-10}$ \\
3 & $-5.024 \times 10^{-10}$ & $1.990 \times 10^{9}$ & $2.876 \times 10^{-14}$ \\
4 & $-1.709 \times 10^{-12}$ & $5.850 \times 10^{11}$ & $6.220 \times 10^{-15}$ \\
5 & $-1.396 \times 10^{-12}$ & $7.166 \times 10^{11}$ & $1.563 \times 10^{-12}$ \\
6 & $-1.230 \times 10^{-13}$ & $8.132 \times 10^{12}$ & $3.473 \times 10^{-14}$ \\
7 & $-2.465 \times 10^{-14}$ & $4.057 \times 10^{13}$ & $8.157 \times 10^{-17}$ \\
8 & $-2.404 \times 10^{-14}$ & $4.159 \times 10^{13}$ & $2.149 \times 10^{-17}$ \\
9 & $-7.904 \times 10^{-17}$ & $1.265 \times 10^{16}$ & $4.080 \times 10^{-20}$ \\ \bottomrule
\end{tabular}
\tablefoot{Modes are sorted by relaxation timescale.}
\end{table}

\begin{table}[b]
\centering
\caption{Extracted parameters for the Kelvin--Voigt elements of the full effective lunar model.}
\label{tab:voigt_elements}
\begin{tabular}{@{}cccc@{}}
\toprule
$i$ & $\tau_i$ (s) & $\alpha_i$ (GPa) & $\eta_i$ (Pa$\cdot$s) \\ \midrule
$1^\ast$ & $4.230 \times 10^{15}$ & $8.273 \times 10^{-4}$ & $3.499 \times 10^{21}$ \\
2 & $4.144 \times 10^{13}$ & $5.232 \times 10^{1}$ & $2.168 \times 10^{24}$ \\
3 & $3.892 \times 10^{13}$ & $1.583 \times 10^{0}$ & $6.161 \times 10^{22}$ \\
$4^\ast$ & $2.624 \times 10^{12}$ & $5.529 \times 10^{-1}$ & $1.451 \times 10^{21}$ \\
5 & $5.854 \times 10^{11}$ & $8.079 \times 10^{3}$ & $4.729 \times 10^{24}$ \\
6 & $1.990 \times 10^{9}$ & $2.479 \times 10^{4}$ & $4.935 \times 10^{22}$ \\
$7^\ast$ & $1.578 \times 10^{6}$ & $1.049 \times 10^{3}$ & $1.655 \times 10^{18}$ \\
$8^\ast$ & $1.058 \times 10^{6}$ & $1.658 \times 10^{3}$ & $1.754 \times 10^{18}$ \\ \bottomrule
\end{tabular}
\tablefoot{Elements retained in the reduced model are marked with an asterisk.}
\end{table}

\begin{table}[t]
\centering
\caption{Love numbers and effective rheological parameters for the lunar homogeneous model.}
\label{tab:effective_parameters}
\resizebox{\columnwidth}{!}{%
\begin{tabular}{@{}lcc@{}}
\toprule
Quantity & Symbol & Value \\
\midrule
Elastic Love number & $k_E$ & $2.40733 \times 10^{-2}$ \\
Fluid Love number & $k_f$ & $1.43752$ \\
Real part of the Love number & $\Rr[k_2]$ & $2.42417 \times 10^{-2}$ \\
Imaginary part of the Love number & $\Ii[k_2]$ & $-5.74359 \times 10^{-4}$ \\
Gravitational rigidity & $\gamma$ & $1.020$ GPa \\
Elastic rigidity & $\alpha$ & $5.987 \times 10^{1}$ GPa \\
Dashpot viscosity & $\eta$ & $6.955 \times 10^{21}$ Pa\,s \\
\bottomrule
\end{tabular}%
}
\end{table}

\begin{figure}[h]
    \centering
    \begin{subfigure}{\columnwidth}
        \centering
        \includegraphics[width=\textwidth]{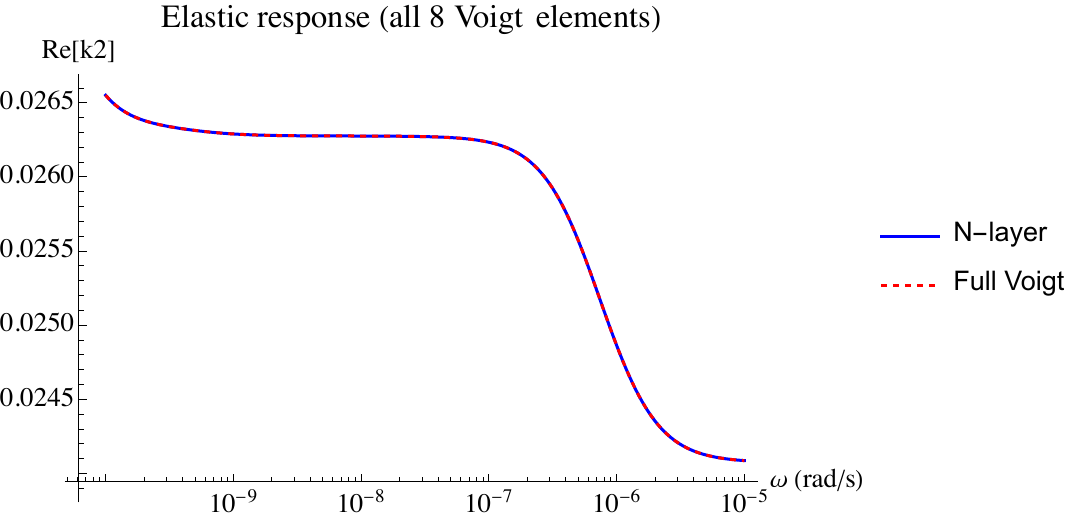}
        \caption{Real part of the Love number, $\Rr[k_2(i\omega)]$.}
        \label{fig:real_full}
    \end{subfigure} \\
    \vspace{0.3cm}
    \begin{subfigure}{\columnwidth}
        \centering
        \includegraphics[width=\textwidth]{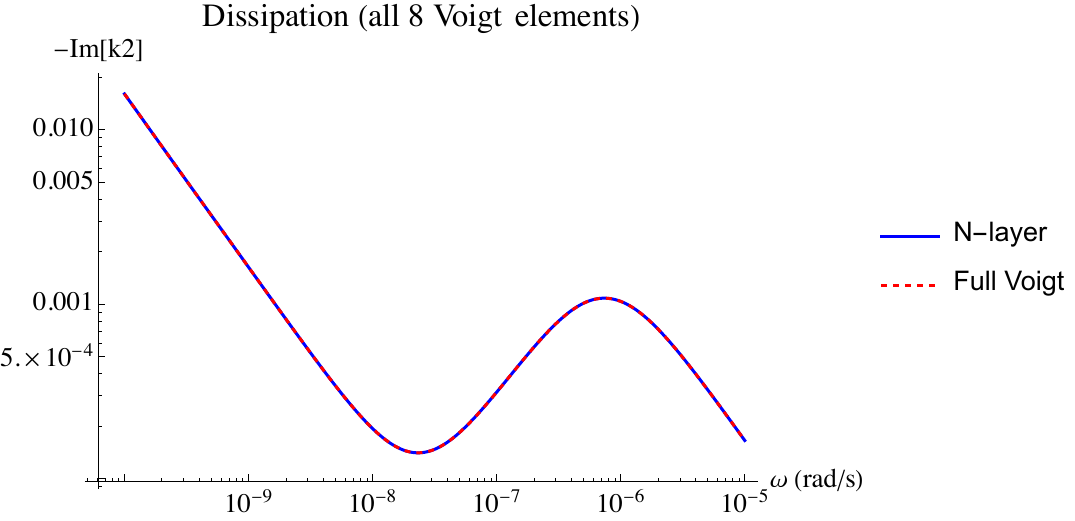}
        \caption{Imaginary part of the Love number, $-\Ii[k_2(i\omega)]$.}
        \label{fig:imag_full}
    \end{subfigure}
    \caption{Comparison between the exact five-layer lunar model and the equivalent homogeneous generalized Voigt model. The real and imaginary parts of the tidal Love number coincide within numerical precision when all extracted Voigt elements are retained.}
    \label{fig:k2_comparison_full}
\end{figure}

In Figure~\ref{fig:k2_comparison_full} we compare the dissipative part $-\Ii[k_2(i\omega)]$, and the elastic part, $\Rr[k_2(i\omega)]$ of the layered moon model to the full homogeneous model. The agreement shows that the full equivalent generalized Voigt model reproduces both the real and imaginary parts of the tidal Love number within numerical precision.

\subsection{Dominant-mode reduction over a frequency interval}

In many applications, it is unnecessary to retain the full effective rheology over the entire frequency range. The package includes a dominant-mode selection step that evaluates the maximum fractional contribution of each Voigt element to $-\Ii[C(i\omega)]$ over a prescribed interval of forcing frequencies and retains only those elements whose contribution exceeds a given chosen threshold.

\begin{figure}[htbp]
    \centering
    \begin{subfigure}{\columnwidth}
        \centering
        \includegraphics[width=\textwidth]{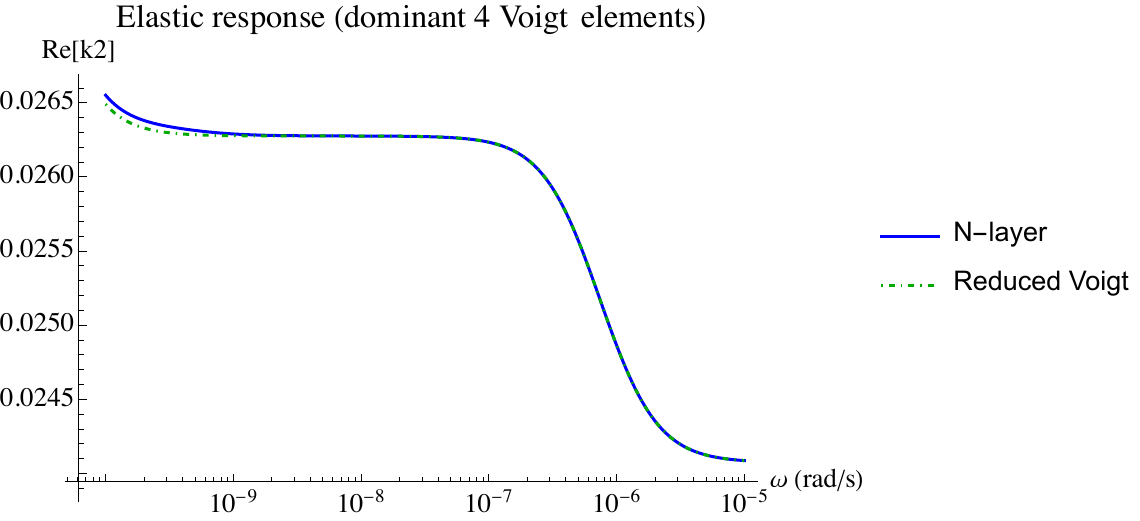}
        \caption{Real part of the Love number, $\Rr[k_2(i\omega)]$.}
        \label{fig:real_reduced}
    \end{subfigure} \\
    \vspace{0.3cm}
    \begin{subfigure}{\columnwidth}
        \centering
        \includegraphics[width=\textwidth]{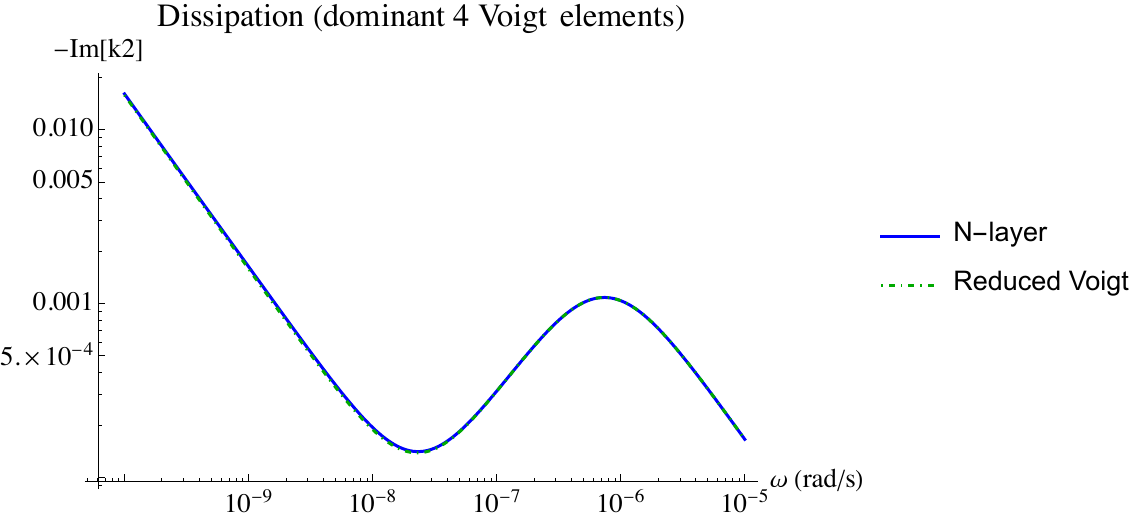}
        \caption{Imaginary part of the Love number, $-\Ii[k_2(i\omega)]$.}
        \label{fig:imag_reduced}
    \end{subfigure}
    \caption{Same as Fig.~\ref{fig:k2_comparison_full}, but for the reduced equivalent homogeneous generalized Voigt model.}
    \label{fig:k2_comparison_reduced}
\end{figure}

For the Moon, the relevant frequency interval is ${\omega \in [10^{-10},\, 10^{-5}]~\mathrm{rad\,s^{-1}}}$ (the default package setting), and we set the dominant-mode threshold to $2\%$. Under these assumptions, the full effective rheology is reduced to the subset of dominant Voigt elements marked with an asterisk in Table~\ref{tab:voigt_elements}. As one can see from Fig.~\ref{fig:k2_comparison_reduced}, the reduced model preserves the main features of the layered model in the selected frequency interval.

The reduction step is important from a practical perspective. It allows the user to tailor the effective homogeneous model to the frequency range relevant for a specific dynamical problem rather than to retain all relaxation modes of the exact inversion. The reduced model is therefore specific to the chosen frequency interval and dominant-mode threshold.

\section{Conclusions}

We have presented a Wolfram Language package that constructs effective homogeneous generalized Voigt rheologies for incompressible, spherically symmetric layered viscoelastic bodies. The package computes the tidal Love number of a multilayer model, extracts its secular poles and residues, and converts this information into the parameters of an equivalent homogeneous rheology. Starting from a user-defined layered model, the package returns the parameters of the corresponding homogeneous generalized Voigt body in a format suitable for validation, comparison, and downstream numerical use and applications.

The lunar case study shows how the code converts a five-layer interior model into an equivalent homogeneous rheology. When all extracted Voigt elements are maintained, the homogeneous model reproduces the layered tidal response within numerical precision. The dominant-mode selection then gives a smaller model adapted to the prescribed frequency interval. The reduction applies when the relevant forcing frequencies lie in a limited range, since it preserves the tidal response of the layered model while reducing the number of rheological elements.

The package allows detailed interior-structure models to be translated into generalized Voigt parameters that can be supplied directly to time-domain tidal evolution codes, enabling their dissipation properties to be incorporated in long-term orbital evolution calculations without repeatedly solving the full layered deformation problem during integration. The same procedure can be applied to other incompressible multilayer models, including models of terrestrial planets and icy satellites, provided that the layer rheologies satisfy the assumptions used in the reduction.

The current implementation assumes a Maxwell rheology for each solid layer. Maxwell bodies underestimate tidal dissipation at forcing frequencies close to or above the inverse Maxwell time, where laboratory and seismological evidence favors rheologies with a broader distribution of relaxation times, such as Andrade or Sundberg--Cooper. Our reduction pipeline is exact for a finite multilayer Maxwell body because the resulting $k_2(s)$ is rational with a finite set of secular poles, a property that rheologies with a continuous relaxation spectrum do not share. Andrade and similar rheologies can nevertheless be approximated within a prescribed frequency range by a finite set of dominant modes, as done in \citet{gev2020} for an extended Burgers model. Extending the package in this direction, together with tighter integration with external dynamical solvers, will require extending the current pole--residue reduction beyond the finite Maxwell case.

\begin{acknowledgements}
The author would like to thank the anonymous referee for careful reading and constructive comments that improved the manuscript. The author would also like to thank Clodoaldo Ragazzo (USP) for many discussions and long-standing collaboration on this topic, Diogo Gomes and Alessandro Astolfi (KAUST) for their kind support, and J.~Ricardo G.~Mendon\c{c}a (USP) for technical advice and a careful reading of the manuscript that greatly improved its presentation. This work was partially supported by the King Abdullah University of Science and Technology (KAUST), Saudi Arabia.
\end{acknowledgements}

\bibliographystyle{aa}
\bibliography{mybibliography}

\end{document}